\documentclass[twocolumn,citeautoscript,superscriptaddress,floatfix,longbibliography]{revtex4-1}

\usepackage[utf8]{inputenc}
\usepackage{amsmath,amssymb,amsmath,amsthm,graphicx,xcolor}
\begin{document}

\author{Stefan Jaschik}
\affiliation{Institut f\"ur Physik, Martin-Luther-Universit\"at
Halle-Wittenberg, 06120 Halle (Saale), Germany}
  
\author{Mário R. G. Marques}
\affiliation{Institut f\"ur Physik, Martin-Luther-Universit\"at
Halle-Wittenberg, 06120 Halle (Saale), Germany}

\author{Michael Seifert}
\affiliation{Institut f\"ur Festk\"orpertheorie und -optik,
  Friedrich-Schiller-Universit\"at Jena, Max-Wien-Platz 1, 07743 Jena, Germany}

\author{Claudia R\"odl}  
\affiliation{Institut f\"ur Festk\"orpertheorie und -optik,
  Friedrich-Schiller-Universit\"at Jena, Max-Wien-Platz 1, 07743 Jena, Germany}
\affiliation{European Theoretical Spectroscopy Facility}
  
\author{Silvana Botti}
\email{silvana.botti@uni-jena.de}
\affiliation{Institut f\"ur Festk\"orpertheorie und -optik,
  Friedrich-Schiller-Universit\"at Jena, Max-Wien-Platz 1, 07743 Jena, Germany}
\affiliation{European Theoretical Spectroscopy Facility}

\author{Miguel A. L. Marques}
\email{miguel.marques@physik.uni-halle.de}
\affiliation{Institut f\"ur Physik, Martin-Luther-Universit\"at
Halle-Wittenberg, 06120 Halle (Saale), Germany}
\affiliation{European Theoretical Spectroscopy Facility}

\date{\today}

\title{Stable ordered phases of cuprous iodide with complexes of copper vacancies}

\begin{abstract}
  We perform an exhaustive theoretical study of the phase diagram of Cu-I binaries, focusing on Cu-poor compositions, relevant for $p$-type transparent conduction. We find
  that the interaction between neighboring Cu vacancies is the determining factor that stabilizes non-stoichiometric zincblende phases. This interaction leads to defect complexes where Cu vacancies align preferentially along the [100] crystallographic direction. It turns out that these defect complexes have an important influence on hole conductivity, as they lead to dispersive conducting $p$-states that extend up to around 0.8~eV above the Fermi level. We furthermore observe a characteristic peak in the density of electronic states, which could provide an experimental signature for this type of defect complexes.
\end{abstract}

\maketitle

Transparent conducting semiconductors (TCSs) form a large family of
materials that combine both high conductivity and transparency. These
are essential properties for the development of transparent electrodes
and thin films transistors in solar cell devices, infrared reflective
coatings, and electrochromic displays, to name a few
examples~\cite{minami2000new,afre2018transparent}. Several $n$-type TCSs
have been found over the past decades, including In$_{2}$O$_{3}$,
SnO$_{2}$, ZnO, and GaN, that are in common use in the industry.

The development of $p$-type TCS materials has been considerably
slower than that of their $n$-type
counterparts~\cite{afre2018transparent}. Since the first report on a $p$-type TCS made of
NiO~\cite{sato1993transparent}, a class of promising $p$-type TCSs was
identified at the end of the 90s in the family of Cu oxides with the
delafossite structure. CuAlO$_{2}$ was first investigated by
Hosono\textit{ et al.} in 1997~\cite{kawazoe1997p}, leading to an
extensive research effort of the whole family of CuMO$_2$ delafossite
compounds~\cite{nagarajan2001ptype} (where M is a trivalent
cation). Until recently, the highest
conductivity of a $p$-type TCS was found in this family, specifically
around 220~S/cm for Mg-doped CuCrO$_2$, although at the cost of low
visible transmittance ( of around 30\% for films with a thickness 250 nm)~\cite{nagarajan2001ptype}. More
recently, a Sr-doped LaCrO$_3$ perovskite was reported as a $p$-type
TCS, exhibiting a moderate conductivity of 54~S/cm and the slightly
better transmittance of 42.3\%~\cite{zhang2015perovskites}.

Only recently, cuprous iodide (CuI), already known to be a transparent semiconductor as early as in 1907~\cite{baedeker1907cuifirst}, has emerged as the most
promising $p$-type TCS~\cite{grundmann2013cuprous}. In this Article we focus on the understanding and control of the exceptional electronic properties of this material.
%In fact, a record
%room-temperature hole conductivity $\sigma > 280$~S/cm with
%transmittances of over 70\% have been achieved for CuI thin
%films~\cite{yang2016room}. 
Recent measurements on CuI have proved optical transparency as high as 90\%~\cite{yamada2016truly} and hole concentrations in the range of
4.0$\times$10$^{16}$--8.6$\times$10$^{19}$~cm$^{-3}$, with
mobilities of 2--43.9~cm$^2$V$^{-1}$s$^{-1}$~\cite{yamada2016truly,grundmann2013cuprous}.
Concerning electron and hole effective masses, values of 0.30(1)~$m_\text{e}$ are reported experimentally for
electrons and 2.4(3)~$m_\text{e}$ for heavy
holes~\cite{honerlage1976carriermass}, while earlier calculations pointed to light holes effective masses in the range
0.2--0.25~$m_\text{e}$~\cite{huang2012first}.  It is known that native Cu vacancies play the role of the dominant
acceptor~\cite{kokubun1971electrical}, and are therefore responsible
for $p$-type conduction of CuI thin films.
The electronic band gap of CuI at $T=80$\,K was measured as
3.1\,eV~\cite{gogolin1989piezobirefringence}, while room temperature experimental
studies report a slightly lower value of
2.93--3.03\,eV~\cite{chaudhuri1990chemical,chen2010growth,ves1981pressure,gruzintsev2012temperature}.
 
Furthermore, there have already been
several successful demonstrations of the use of CuI in opto-electronic
devices: as a hole transport layer in solid-state
dye-sensitized~\cite{amalina2013properties} and perovskite solar
cells~\cite{christians2014an,sepalage2015copper,yu2015recent}, as a
hole-selective contact in organic solar cells~\cite{mohamed2015cui},
or in light emitting diodes~\cite{shan2017enhanced}. Recently, it was
used as a transparent flexible thermoelectric
material~\cite{yang2017transparent}.

CuI has a series of advantages. It is an environment-friendly
material composed of non-toxic and naturally abundant elements.  The
zincblende structure is easy to match with other conventional
semiconductors and the direct band gap of about 3~eV is also beneficial
for the preparation of the $p$–$n$ junctions~\cite{godinho2010understanding}. From the chemistry point of view it
is a rather interesting and unique compound. It is stable in the
Cu$^\text{I}$ oxidation state, while in other halide salts of copper
the Cu$^\text{II}$ oxidation state is favored. In part this is due to the very large difference
in ionic radii between Cu and I ($r_{\text{I}^-} =
206$\,pm~\cite{shanon1976revised}), difference that increases as we
move from Cu$^\text{I}$ and Cu$^\text{II}$, thereby decreasing
stability. Moreover, I$^{-}$ is a powerful reducing agent, capable of
reducing Cu$^\text{II}$ to Cu$^\text{I}$ spontaneously. In spite of
this, it is actually possible to form cupric iodide, CuI$_2$, by the
interaction between cuprous iodide and iodine solution. This leads to a
greenish-blue solution that unfortunately could not be isolated in the
solid state.

We find a rather diverse structural variety in CuI, as it can exist in
several polymorphs. At ambient conditions it crystallizes in the
zincblende structure (the so-called $\gamma$-phase), while there are
two high-temperature phases, the cubic $\alpha$ and the wurtzite
$\beta$ polymorphs~\cite{miyake1952on,grundmann2013cuprous}. Note
that there also exist a couple of trigonal phases~\cite{kurdyumova1961cui,sakuma1988crystal,villars2009cui,akopyan2010specific} that are layered, with a bonding pattern rather different from the $\gamma$-phase.

It is tempting to draw a comparison between $\gamma$-CuI and another
Cu compound relevant for solar cell applications, specifically
Cu(In,Ga)(S,Se)$_2$ (CIGS) and Cu$_2$ZnSn(S,Se)$_4$ (CZTS). In
these compounds we have a similar crystal structure, that includes a Cu$^\text{I}$
cation (crystallographically, the kesterite and the chalcopyrite structures
are closely related to zincblende) and a direct band gap. Furthermore, in both cases, native Cu vacancies have a very low formation energy, and are present in large
quantities~\cite{baranowski2016review}. In fact, CIGS solar cells can
be produced with efficiencies above 20\% with Cu/(Ga+In) ratios from
0.80 to 0.92~\cite{jackson2011new}. It is also known that in CIGS
neutral defect complexes are energetically favorable,
particularly the complex formed by two Cu vacancies and a
Cu-\{In,Ga\} anti-site~\cite{guillemoles2000stability}. This leads to the
formation of ordered-defect phases, where an attractive
interaction among the defect pairs leads to the observed
stoichiometries CuIn$_5$Se$_8$, CuIn$_3$Se$_5$, Cu$_2$In$_4$Se$_7$,
etc.~\cite{monig2010surface,schmid1993chalcopyrite,zhao2012density,zhang1997stabilization}.

Going back to CuI, the record
room-temperature hole conductivity $\sigma > 280$~S/cm, with
transmittance of over 70\%, has been measured for CuI thin
films~\cite{yang2016room}. 
This was achieved for an iodine-rich growth condition, which creates Cu vacancies and is therefore advantageous for inducing holes in CuI~\cite{maurer1945deviations}. On the other hand, it has been shown that, among the CuI native defects, Cu vacancies have the lowest formation and ionization energies for both Cu-rich or I-rich equilibrium growth conditions~\cite{wang2011native,migle2019}.

While the importance of the role played by Cu vacancies for $p$-type transparent conductivity is fully acknowledged, it remains unclear how these Cu vacancies arrange themselves in CuI and what is their optimal and maximum concentration. In particular, the formation of ordered-defect compounds, as it happens in similar CIGS semiconducting absorbers, has not been investigated yet. The aim of this work is to provide an answer to these unresolved questions.

\begin{figure}[th]
  \begin{center}
  \includegraphics[width=8cm]{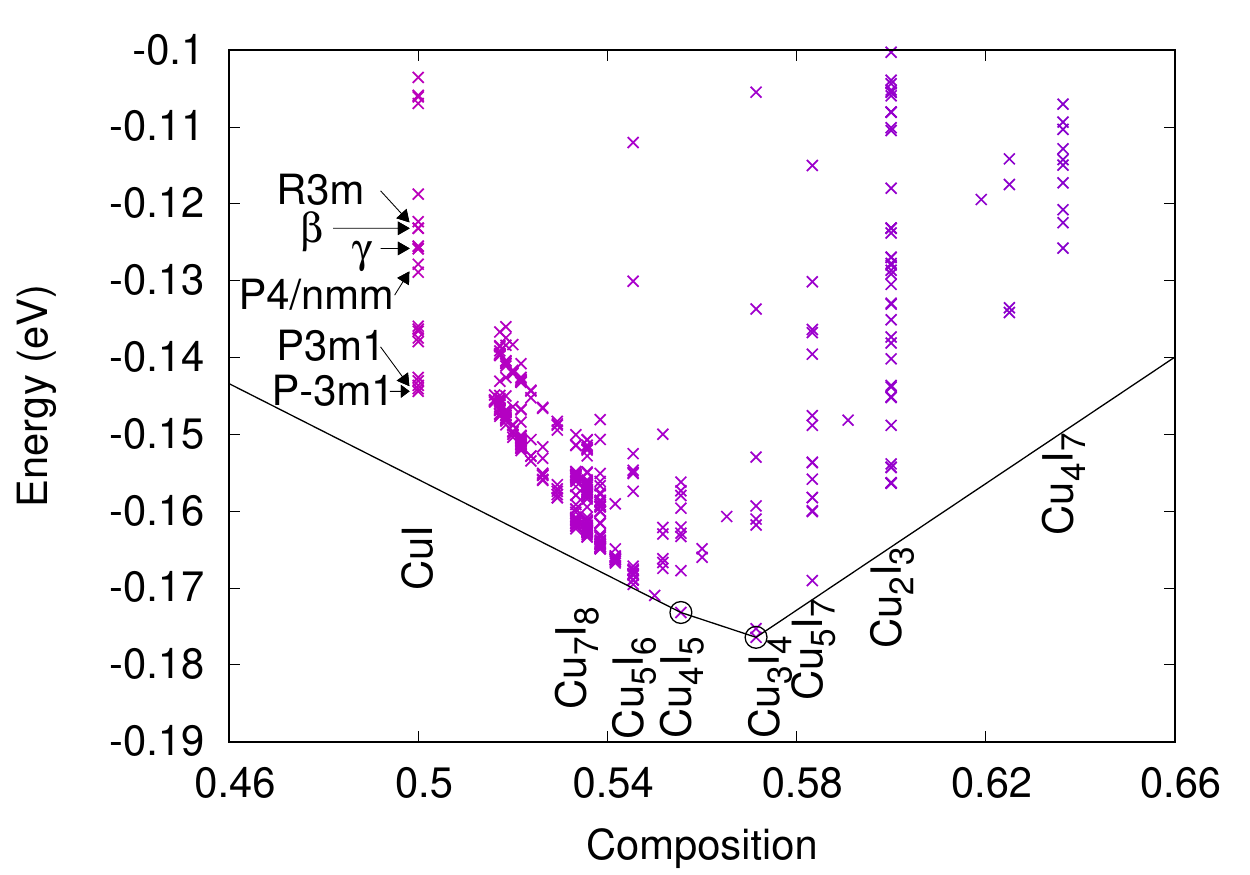} \\
  \end{center}
  \caption{Binary phase diagram of Cu$_{1-x}$I$_{x}$ obtained with the SCAN functional~\cite{sun2015strongly}. We only show the relevant phases for $p$-type transparent conduction
    ($x\in[0.45,0.65]$). Points that are strictly on the convex hull
    of thermodynamic stability (specifically, the lowest energy
    crystal structures Cu$_4$I$_5$ and Cu$_3$I$_4$) are represented by
    circles, while other phases are indicated by crosses. The chemical
    potentials of the elementary phases are set to zero, so the energy
    values indicated are in fact formation energies. The space groups identify some structures found in the materials project~\cite{materialsproject}, and $\gamma$ and $\beta$ indicates the zincblende and wurzite structures, respectively. }
 \label{graf:phadiag}
 \end{figure}

We study therefore the stability of Cu-I compounds, focusing on the interaction between Cu vacancies in
$\gamma$-CuI, and if this interaction can lead to the formation of
ordered-defect compounds. Our strategy is to perform a rather
exhaustive computational exploration of the phase diagram at zero temperature of the binary Cu--I system, 
focusing on the Cu-poor region, up to 30\% concentration of Cu vacancies.

We begin our exploration by using global structural prediction
methods~\cite{OganovBook}. These are completely unbiased techniques
that allow, for a given composition, the prediction of the
ground-state structure, and of some of the low-energy
polymorphs. Specifically, we applied the minima-hopping method
(MHM)~\cite{goedecker2004minima,amsler2010crystal}, an efficient
approach that has been frequently used in the past years to discover
new materials and crystal
phases~\cite{borlido2017structural,shi2017high}.  We recall that this method uses a walker in configuration space, and that it proceeds by a series of short molecular dynamics simulations followed by geometry
optimization steps.  We start the MHM with a random structure, an initial temperature of 500\,K, and a variable time-step for the MD simulations. The energies and forces required for the
simulations were obtained at the level of density-functional theory
(DFT) with the Perdew-Burke-Ernzerhof~\cite{perdew1996generalized}
(PBE) approximation to the exchange-correlation functional as
implemented in the {\sc vasp}
code~\cite{kresse1996efficient,kresse1996efficiency}. We used the PAW
pseudopotentials of version 5.2 of {\sc vasp} (11 valence electrons for Cu and 7 for I), an energy cut-off of 520 eV, and a constant density
of  1000 per atom k-points that yields a precision of around 2~meV/atom
in the total energy. This amounts to a 8x8x8 mesh for the primitive
cell of zincblende CuI.

With the MHM we explored the complete phase diagram of Cu$_{1-x}$I$_x$, with 0$<x<$1.
For efficiency reasons, we restricted this step of the study to a rather small number of atoms in the
unit cell (maximum 6), and we stopped our minima
hopping runs relatively early (after $\sim$30--80 minima were
found). We note, however, that our main objective was to get an
overall view of the phase diagram, for which these simulations proved
sufficient. From our results it was rather clear that for the
region of the phase diagram we are most interested in, ranging from
CuI to about Cu$_2$I$_3$ (corresponding to 33\% of Cu vacancies), all
lowest energy phases corresponded to defected zincblende CuI.
On the other hand, a very large concentration of Cu vacancies inevitably leads to a breakdown of the parent zincblende structure.

The next step was to perform an extensive calculation of all possible
crystal structures of $\gamma$-CuI including a varying number of Cu
vacancies, focusing on the Cu poor region of the phase diagram up to a Cu vacancy concentration of about 30\%. For that, we used software included in {\sc
  atat}~\cite{walle2009multicomponent} to generate all possible
supercells containing up to 14 sites (7 Cu atoms) and up to 50\%
Cu vacancies (i.e. $1/2<x<2/3$). The geometries of these 118 unique configurations were
then optimized within DFT and the PBE approximation, and added to the
phase diagram.

Unfortunately, and due to the combinatorial nature of the problem, it
was not possible to systematically study supercells with a larger
number of atoms with DFT. As a workaround, we fitted a cluster
expansion with {\sc atat}~\cite{walle2009multicomponent} to the
results for the small unit cells, and used it to predict the formation
energy of all possible supercells containing up to 32 sites (16 Cu
atoms). The cluster expansion was capable of a very good fit, yielding
a cross-variation error of 33~meV/atom. From all these 30\,849
geometries, we then selected the ones that were close to the convex
hull, that were then reoptimized with {\sc vasp} and added to the
phase diagram.

It is well known that the PBE approximation, a generalized gradient approximation (GGA), leads to rather large
errors for the calculation of formation energies, on average more than
200~meV/atom~\cite{perez2015optimized,tran2016rungs,stevanovic2012correcting}. Therefore, in
order to increase the reliability of our results we decided to
reoptimize all structures with the strongly constrained and
appropriately normed~\cite{sun2015strongly} (SCAN) functional.
This is a meta-GGA, designed to obey 17 exact constrains of the
exchange-correlation functional, albeit at the cost of larger
complexity with respect to the PBE. In terms of results, it represents
an remarkable improvement in formation energies with respect to
PBE~\cite{Isaacs2018performance,zhang2018efficient}, cutting the average error by
more than a factor of two for main group compounds.

\begin{figure}[th!]
  \begin{center}
    \includegraphics[width=7cm]{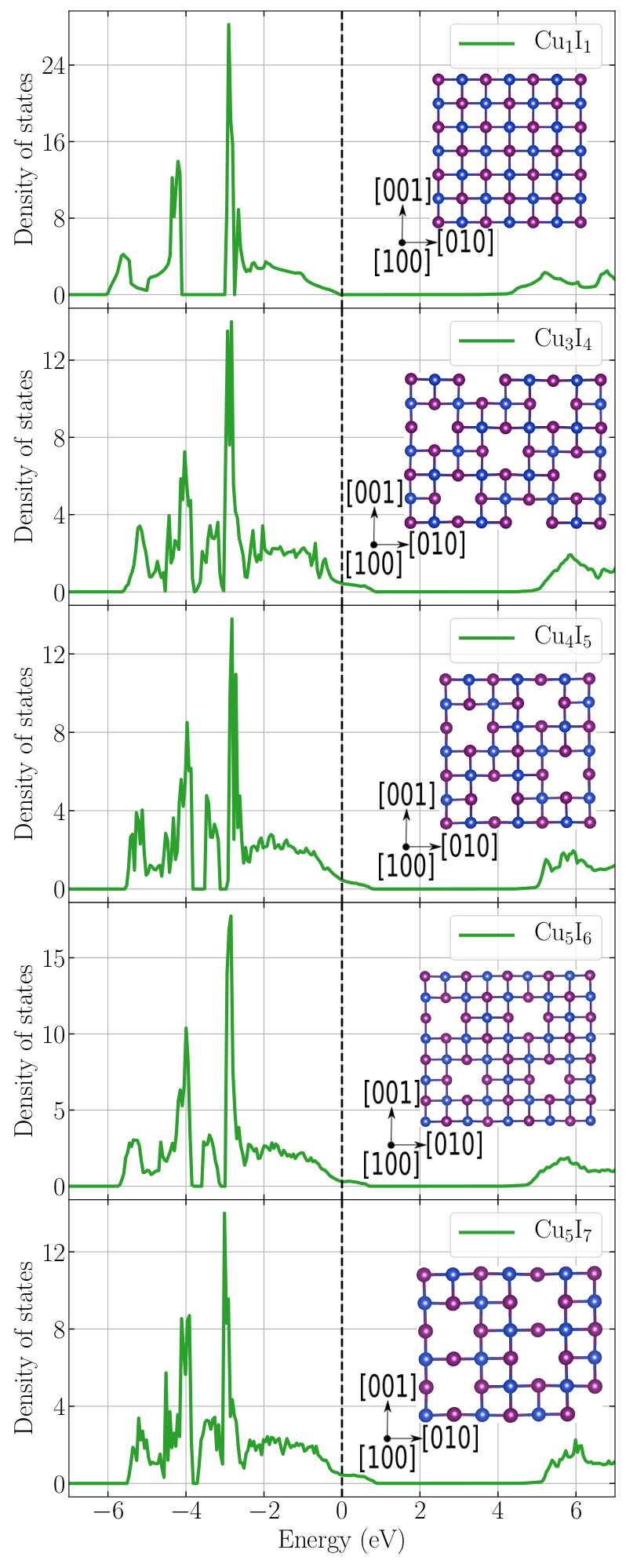} \\
  \end{center}
  \caption{Density of electronic states for the lowest energy
    configuration of Cu$_1$I$_1$, Cu$_3$I$_4$, Cu$_4$I$_5$,
    Cu$_5$I$_6$, and Cu$_5$I$_7$, which are shown in the inset. Cu
    atoms are in blue, while I atoms are in purple. The lines of
    Cu vacancies run along the crystallographic [100] axis of the
    original zincblende structure. The insets were produced with {\sc
      vesta}~\cite{momma2011VESTA}.  The curves were calculated with
    the PBE0 hybrid exchange-correlation
    functional~\cite{adamo1999toward,ernzerhof1999assessment}, and
    were normalized to the number of I atoms in each structure.}
 \label{graf:dos}
\end{figure}

In total, we computed 623 unique Cu$_{1-x}$I$_{x}$ phases, whose SCAN
energy is plotted in Fig.~\ref{graf:phadiag}. The ordered phase with lowest
formation energy is Cu$_3$I$_4$, followed by Cu$_4$I$_5$. However, the
whole region in between these two compositions is very close to the
convex hull. Decreasing the concentration of Cu vacancies, we see that
the formation energies increase in a rather smooth manner,
approaching, as expected, the energy per atom of $\gamma$-CuI. We remark that
the inclusion of configurational entropy in the calculation of the free energy would
further favour configurations with a high concentration of Cu vacancies.

The geometries of some of the structures with energies close to the
convex hull are depicted as insets in Fig.~\ref{graf:dos}. All structures were reoptimized at the end using SCAN. The crystallographic information is available as Supporting Information.
We can see that, beside an overall contraction of the structure, there is no evident atomic rearrangement upon relaxation.
We can always observe the same pattern: Cu vacancies order in
lines along the crystallographic [100] direction. The difference
between the different structures regards how these vacancy lines are
organized with respect to each other. This is, in our opinion, a
remarkable result that indicates that the physics of $p$-type CuI may
not be, in fact, dictated by isolated vacancies, but by complex
defects where the vacancies are to some extent ordered.

Although the lowest-energy structures for different compositions exhibit consistently lines of vacancies in the [100] direction, we could also find low-energy complexes with different patterns, lying only a few meV/atom higher in energy. On one hand, this observation reinforces our conclusion that strong vacancy interactions stabilize Cu-I binary systems. On the other hand, we must expect to have disordered vacancy configurations in real samples at finite temperature, due to entropic effects.

We want anyway to consider the question if it is possible to synthesize
experimentally any of these ordered-defect phases, similarly to what
happens in
CIGS~\cite{monig2010surface,schmid1993chalcopyrite,zhao2012density,zhang1997stabilization}. This
might be true for compositions like Cu$_4$I$_5$ or Cu$_3$I$_4$, where
we find a large gap between the ground-state structure and the other
polymorphs, but not for smaller concentration of Cu vacancies. In
fact, for such compositions we find a large number of polymorphs very
close in energy (within a few meV/atom). Again, the main difference
between these structures is how the lines of vacancies are
distributed, which is likely to lead to some disorder in experimental
samples. 
We confirmed the dynamical stability of the most promising ordered compounds, Cu$_3$I$_4$ and Cu$_4$I$_5$, by calculating their phonon band structure and comparing them with the one of $\gamma$-CuI. The phonon band structures are available as Supporting Information.

Concerning stoichiometric CuI, we find, in agreement
with other theoretical works~\cite{jain2013commentary}, that the layered structure of
CuI has the lowest energy of all polymorphs, 18~meV/atom below the
zincblende phase with the SCAN functional. Note that this difference
is considerably higher than with the PBE, where it stands at a mere
2~meV/atom. This stoichiometry does not seem to be
thermodynamically stable, as it lies 11~meV/atom above the convex
hull. However, we can expect that van der Waals interactions, that are absent from SCAN,
stabilize the layered compound with respect to the (defected) zincblende geometries.

We now turn to the study of the electronic properties of our
structures. Due to the systematic underestimation of the band gap and
incorrect positioning of the $d$-states of semi-local
exchange-correlation functionals (like the PBE or SCAN), we decided to
use a hybrid functional. We chose for our calculations the hybrid
based on the PBE, specifically
PBE0~\cite{adamo1999toward,ernzerhof1999assessment}, as it gives an
excellent description of the band gap of zincblende
CuI~\cite{koyasu2019optical}.

From the PBE0 band structure we fitted light-hole effective masses in the $\Gamma - X$ direction of 0.31~$m_\text{e}$ (0.22~$m_\text{e}$ with PBE).  Note that the band dispersions are very similar in PBE and PBE0, despite the big difference in the size of the band gap.

We further studied the electronic band structure and the optical absorption spectrum of Cu$_3$I$_4$, as this compound has a not too big unit cell, and it is also the structure with the highest number of vacancies (and therefore the lowest chemical potential in the valence band). The effective masses remain small, which is an indication of high mobility of holes. We obtained in fact $0.32$~$m_\text{e}$ for the light-hole effective mass along  $\Gamma - X$ using PBE. The absorption spectra of $\gamma$-CuI and Cu$_3$I$_4$ are available as Supporting Information. Due to the small effective mass, the valence band of Cu$_3$I$_4$ is very dispersive. The Fermi energy of Cu$_3$I$_4$ lies therefore almost 1~eV below the valence band maximum. This leads to a Drude peak in the absorption spectrum, whose tail extends to the visible range. By contrast, zincblende CuI is transparent up to 3.1~eV. For a use of this material as a transparent conductor, one has therefore to reach a compromise between the thickness of the TCS layer (that will determine light absorption) and the number of vacancies (that will determine the concentration of charge carriers and the position of the Fermi energy in the gap). 

We suggest two strategies to suppress the Drude tail. One possibility is to push up the Fermi energy by doping with compensating impurities. Some of the authors of this Article have recently performed a high-throughput study of doping in CuI~\cite{migle2019}, showing that Mg, Be, or Sn impurities on empty Cu sites would be particularly suitable for this purpose. 
Another possibility is to use very thin CuI layers, profiting from the very large number of holes in the ordered Cu$_3$I$_4$ phase. A good transparency may then be attained in spite of the Drude tail.

\begin{figure*}[th]
  \begin{center}
    \begin{tabular}{c@{\hskip 0.5cm}c@{\hskip 0.5cm}c}
      \includegraphics[height=3cm]{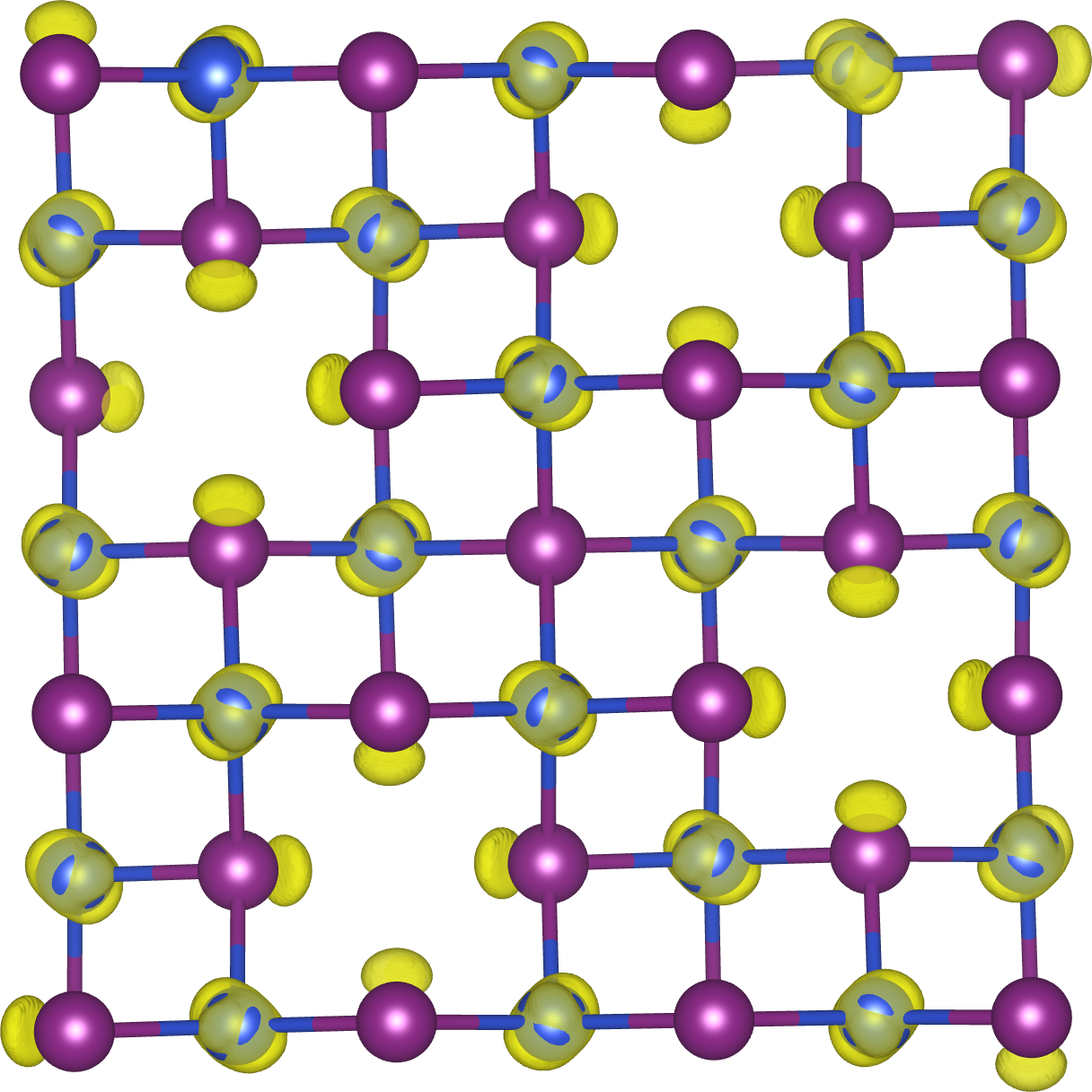} &
      \includegraphics[height=3cm]{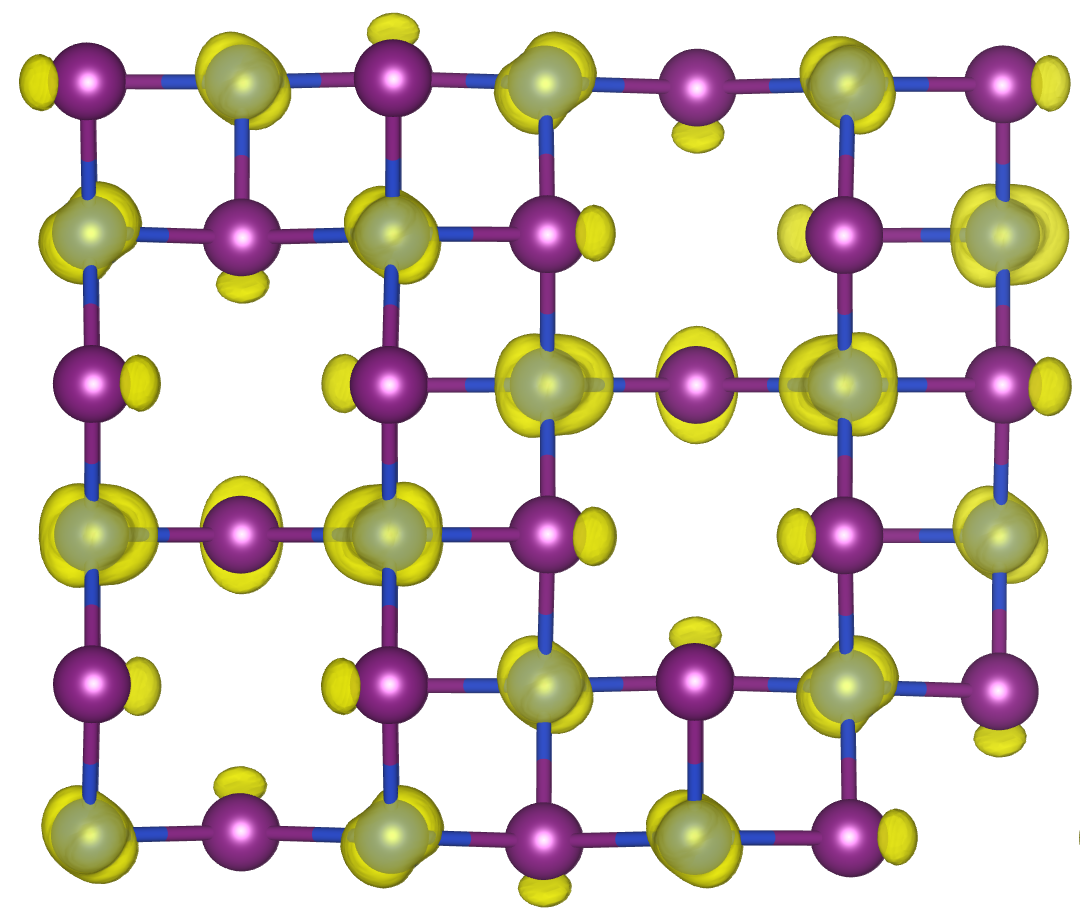} &
      \includegraphics[height=3cm]{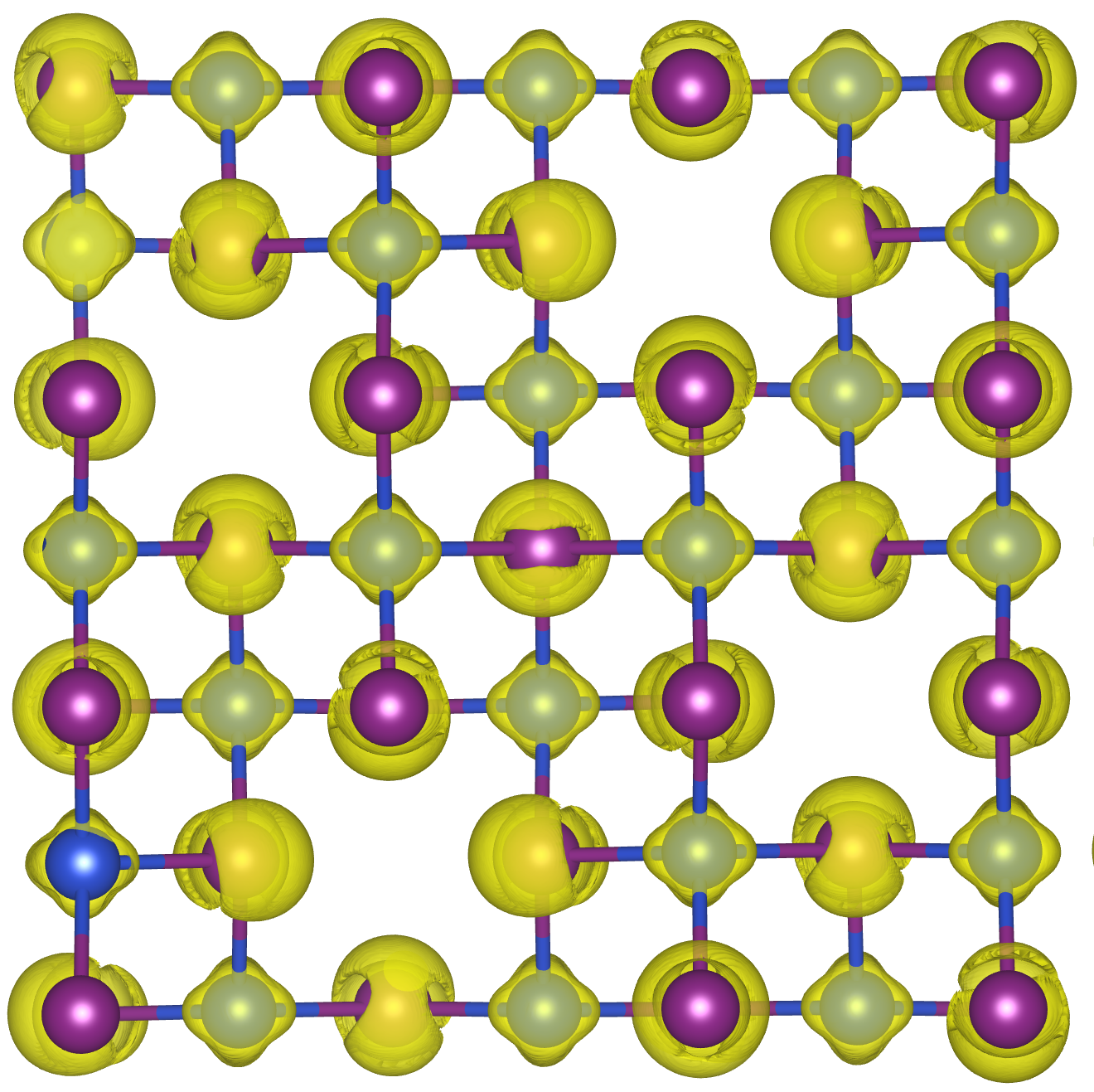} \\
      Cu$_4$I$_5$ & Cu$_5$I$_7$ & Cu$_4$I$_5$ \\
%      [-2.77,-2.47] & [-2.91,-2.06] & [-0.01,0.43]  \\
    \end{tabular}
  \end{center}
  \caption{Partial charge density (yellow) integrated for the states
    around -3~eV for Cu$_4$I$_5$ (left) and Cu$_5$I$_7$ (middle) and
    for the conduction states of Cu$_4$I$_5$ (right) above the Fermi energy.
    This image was produced with {\sc vesta}~\cite{momma2011VESTA}.}
 \label{graf:los}
 \end{figure*}
% Cu4I5 / Ef=3.91653464 / [1.15,1.45] - 0.023 e/a_0^3 / [3.91,4.35] - 0.0019 e/a_0^3
% Cu5I7 / Ef=3.66284307 / [0.75,1.60] - 0.023 r/a_0^3

The density of electronic states (DOS) of the most relevant low-energy
crystal structures are compared to the one of pristine zincblende CuI
in Fig.~\ref{graf:dos}. All curves are aligned to their respective
Fermi energies. We can see that the DOS for all defect structures are
very similar. The band gap remains similar to the one of pristine CuI,
and we find hole conduction states extending from the Fermi energy up
to around 0.7--0.9~eV, consisting mainly of I $p$-states. This is
well below the visible range, in agreement with experiment that finds
that $p$-type CuI is indeed transparent in this range. An example of
such hole states is presented in Fig.~\ref{graf:los} (right panel) for
Cu$_4$I$_5$ (the states are very similar for the other compositions of
Fig.~\ref{graf:dos}). It is evident that these states are rather
delocalized in the whole Cu--I framework, which is to be expected from
the small hole mass of the CuI $p$-type conduction states.

Finally, we see a new peak emerge between -3.7 and -3~eV, that is
composed by Cu $d$-states hybridized with I $p$-states. This is, in
our opinion, an important feature of the DOS, as it may provide a
spectroscopic signature of the defect complexes. These states are
depicted, in Fig.~\ref{graf:los}, for both Cu$_4$I$_5$ (left panel)
and Cu$_5$I$_7$ (middle panel). From the picture it is clear that the
contribution coming from the $p$-states of I is localized inside the lines of Cu
vacancies.

In conclusion, we find that the interaction between Cu vacancies
has a stabilizing effect in zincblende CuI. In crystalline compounds 
vacancies organize in lines
along the [100] crystallographic direction. Moreover, CuI is
predisposed to admit a large range of concentration of Cu vacancies, with binary compounds on the convex hull of stability, or very close to it, with a 10--30\% content of Cu vacancies. Our calculations indicate that it should be possible for phases like
 Cu$_4$I$_5$ and Cu$_3$I$_4$, to form ordered defect compounds.
We note that defect complexes are stabilized due to
energetic, and not to entropic effects, which are not included in our calculations. Entropic effects
are further expected to lower the free energy of disordered arrangements of vacancies. 
We believe therefore that the physics of $p$-type zincblende CuI should be
discussed based on defect complexes and not only on the basis of
isolated Cu vacancies.

These complex defects give rise to $p$-type conduction states that
extend for 0.7--0.9~eV above the Fermi level. 
Taking advantage of the empty Cu sites, compensating doping by donor impurities can be used to tune the position of the Fermi energy,
not to hinder transparency in the visible range. Moreover, we find a
clear splitting of the deeper lying electronic states, which may
provide an experimental signature for the existence of lines of
Cu vacancies.

\section*{Acknowledgments}

M.A.L.M. and S.B. acknowledge financial support from the Deutsche
Forschungsgemeinschaft (DFG, German Research Foundation) through the
projects SFB-762 (project A11), SFB-1375 (project A02), BO 4280/8-1 and
BO 4280/8. C.\,R.\ acknowledges financial support from the Marie
Sk\l{}odowska-Curie Actions (GA No.~751823). Crystal structures were
visualized with {\sc vesta}~\cite{momma2011VESTA}.

\bibliography{bib_file}

\section*{Table of contents (ToC image)}
\begin{figure*}[th!]
  \begin{center}
    \includegraphics[width=7cm]{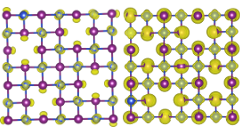} \\
    $P$-type conduction states of ordered-defect CuI compounds.
  \end{center}
 \label{graf:toc}
\end{figure*}

\end{document}